\documentclass[twocolumn,prl,showpacs,superscriptaddress,preprintnumbers,amsmath,amssymb]{revtex4}
\usepackage{graphicx}% Include figure files
\usepackage{dcolumn}% Align table columns on decimal point
\usepackage{bm}% bold math
\usepackage{color}

\begin{document}

\newcommand{\siox}{SiO$_2$}
\newcommand{\silicate}{Si$_2$O$_3$}
\newcommand{\sicoxint}{\mbox{SiC/SiO$_2$}}
\newcommand{\rootthree}{($\sqrt{3}$$\times$$\sqrt{3}$)R30$^{\circ}$}
\newcommand{\rootthreehl}{($\sqrt{\mathbf{3}}$$\times$$\sqrt{\mathbf{3}}$)R30$^{\circ}$}
\newcommand{\sixroot}{(6$\sqrt{3}$$\times$6$\sqrt{3}$)R30$^{\circ}$}
\newcommand{\sixroothl}{(6$\sqrt{\mathbf{3}}$$\times$6$\sqrt{\mathbf{3}}$)R30$^{\circ}$}
\newcommand{\three}{\mbox{(3${\times}$3)}}
\newcommand{\four}{\mbox{(4${\times}$4)}}
\newcommand{\five}{\mbox{(5${\times}$5)}}
\newcommand{\six}{\mbox{(6${\times}$6)}}
\newcommand{\two}{\mbox{(2${\times}$2)}}
\newcommand{\twobyc}{\mbox{(2$\times$2)$_\mathrm{C}$}}
\newcommand{\twobysi}{\mbox{(2$\times$2)$_\mathrm{Si}$}}
\newcommand{\seven}{($\sqrt{7}$$\times$$\sqrt{7}$)R19.1$^{\circ}$}
\newcommand{\fourseven}{(4$\sqrt{7}$$\times$4$\sqrt{7}$)R19.1$^{\circ}$}
\newcommand{\one}{\mbox{(1${\times}$1)}}
\newcommand{\sicbar}{SiC(000$\bar{1}$)}
\newcommand{\sicbarhl}{SiC(000$\bar{\mathbf{1}}$)}
\newcommand{\bardir}{(000$\bar{1}$)}
\newcommand{\grad}{\mbox{$^{\circ}$}}
\newcommand{\cgrad}{\,$^{\circ}$C}
\newcommand{\projecta}{(11$\bar{2}$0)}
\newcommand{\projectb}{(10$\bar{1}$0}
\newcommand{\projectc}{(01$\bar{1}$0}
\newcommand{\third}{$\frac{1}{3}$}
\newcommand{\thirdspot}{($\frac{1}{3}$,$\frac{1}{3}$)}
\newcommand{\oversix}{$\frac{1}{6}$}
\newcommand{\kpoint}{$\bar{\textrm{K}}$-point}
\newcommand{\kpar}{\underline{k}$_{\parallel}$}

%\preprint{APS/123-QED}

%\title{Decoupling Epitaxial Graphene from SiC via reversible Hydrogen Intercalation}

\title{Quasi-free Standing Epitaxial Graphene on SiC by Hydrogen Intercalation}

\author{C. Riedl}
\affiliation{Max-Planck-Institut f\"{u}r Festk\"{o}rperforschung,
Heisenbergstr. 1, D-70569 Stuttgart}

\author{C. Coletti}
\affiliation{Max-Planck-Institut f\"{u}r Festk\"{o}rperforschung,
Heisenbergstr. 1, D-70569 Stuttgart}

\author{T. Iwasaki}
\affiliation{Max-Planck-Institut f\"{u}r Festk\"{o}rperforschung,
Heisenbergstr. 1, D-70569 Stuttgart}

\author{A. A. Zakharov}
\affiliation{MAX-Lab, Lund University, Box 118, Lund, S-22100, Sweden}%

\author{U. Starke}%
 \email{u.starke@fkf.mpg.de}
 \homepage{http://www.fkf.mpg.de/ga}
\affiliation{Max-Planck-Institut f\"{u}r Festk\"{o}rperforschung, Heisenbergstr. 1, D-70569 Stuttgart}

\date{\today}

\begin{abstract}
Quasi-free standing epitaxial graphene is obtained on SiC(0001) by
hydrogen intercalation. The hydrogen moves between the
{(6$\sqrt{3}$$\times$6$\sqrt{3}$)R30$^{\circ}$} reconstructed
initial carbon layer and the SiC substrate. The topmost Si atoms
which for epitaxial graphene are covalently bound to this buffer
layer, are now saturated by hydrogen bonds. The buffer layer is
turned into a quasi-free standing graphene monolayer with its
typical linear {$\pi$}-bands. Similarly, epitaxial monolayer
graphene turns into a decoupled bilayer. The intercalation is stable
in air and can be reversed by annealing to around
900{\,$^{\circ}$C}.
% 597 characters

\end{abstract}

%\pacs{Valid PACS appear here}

\maketitle

% main text

\begin{figure}[h!]
\begin{center}
\includegraphics[width=0.47\textwidth]{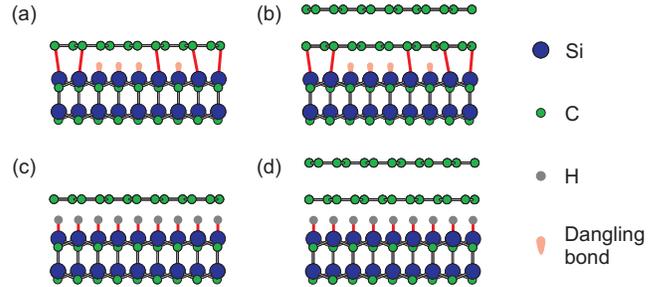}
\end{center}
\caption{Side view models for (a) the {\sixroot}
reconstruction of SiC(0001) ("zerolayer") and (b) epitaxial
monolayer graphene. After hydrogen intercalation (c) the zerolayer
and (d) monolayer graphene are decoupled from the substrate.}
\label{models}
\end{figure}

Graphene, a mono-atomic layer of graphite, displays outstanding
electronic, optical, mechanical and thermal properties which make it
extremely appealing for a wide range of applications~\cite{Geim2009,
Berger2006}. Grown on hexagonal silicon carbide (SiC) wafers, large
area epitaxial graphene samples appear feasible and integration in
existing device technology can be
envisioned~\cite{Berger2006,Ohta2006,Riedl2008}. And indeed, the
achievement of large scale homogeneity is on the
way~\cite{Emtsev2009}. While epitaxial graphene on {\sicbar} allows
only for a poor thickness control and the graphene grows with
rotational disorder \cite{Berger2006,Hiebel2008,Starke2009}, on
SiC(0001) a defined number of epitaxially ordered graphene layers
can be grown~\cite{Riedl2008,Emtsev2009,Riedl2007}. However, an
intrinsic electron doping (n $\approx$ $10^{13} \textrm{cm}^{-2}$)
is observed~\cite{Ohta2006,Riedl2008} which originates from the
influence of the {\sixroot} reconstructed interface
layer~\cite{Mallet2007,Riedl2007} present between graphene and SiC.
This interface or buffer layer is constituted of carbon atoms which
are arranged in a graphene-like honeycomb structure. However,
as depicted in the model sketch in Fig. \ref{models}
(a), about 30\% of these carbon atoms are bound to the Si atoms of
the SiC(0001) surface \cite{Emtsev2008,Mattausch2007}, which
prevents {$\pi$}-bands with the linear dispersion typical for
graphene to develop in this layer. Thus, the interface layer is
electronically inactive in terms of the typical graphene properties
so that it is often called zerolayer graphene. The
second carbon layer grows on top of the interface without covalent
interlayer bonds as shown in Fig. \ref{models} (b) and
electronically acts like monolayer graphene. The influence of
the covalent bonding in the interface layer is also
one of the primary suspects for the strongly reduced mobility in
epitaxial graphene on SiC(0001) as compared to exfoliated graphene
flakes, probably due to the introduction of scattering centers into
the graphene layer. So, for a practical application of epitaxial
graphene on SiC(0001) it would be desirable to counteract the
intrinsic doping and to reduce the influence of the interface
bonding to create quasi-free standing layers.

Reduction of the intrinsic charge carrier density by surface
transfer doping has been achieved recently by means of deposited
F4-TCNQ molecules~\cite{Coletti2009} or atomic layers of Bi and
Sb~\cite{Gierz2008}. Elimination of the covalent bonding at the
interface in order to decouple the epitaxial graphene layers from
the SiC substrate would require to break and saturate the respective
bonds. As sketched for zero- and monolayer graphene
in Fig. \ref{models} (c) and (d) we demonstrate in the present
letter that hydrogen intercalation can induce the desired
decoupling. As a result the outstanding properties
of graphene can be made accessible in quasi-free standing epitaxial
graphene layers on large-scale SiC(0001) wafers suitable for a
practical technological application.

For our experiments on-axis oriented 4H- and 6H-SiC(0001) samples
doped with nitrogen (10$^{17}$ to 10$^{18}$\,cm$^{-3}$ range) were
prepared by chemical-mechanical polishing or hydrogen
etching~\cite{Soubatch2005,Frewin2009} in order to get a regular
array of atomically flat terraces. The epitaxial graphene layers
were prepared by graphitization under ultrahigh vacuum conditions
\cite{Ohta2006,Riedl2008,Riedl2007} or in an induction furnace under
Ar atmosphere \cite{Emtsev2009}. After transport through air the
samples were annealed at temperatures between 600{\cgrad} and
1000{\cgrad} in molecular hydrogen at atmospheric pressures. The
process was carried out in a chemical vapor deposition reactor in an
atmosphere of palladium-purified ultra-pure molecular hydrogen,
similar to the technique used for hydrogen
etching~\cite{Soubatch2005,Frewin2009} and hydrogen
passivation~\cite{Tsuchida1999, Seyller2004, Coletti2008} of SiC
surfaces. Angle-resolved photoelectron spectroscopy (ARPES) using
monochromatized He II radiation, low energy electron diffraction and
microscopy (LEED, LEEM) and core level photoemission spectroscopy
(CLPES) were used to analyze the structural, electronic and
morphological properties of the epitaxial graphene layers after
hydrogen intercalation and the effect of subsequent annealing. The
CLPES experiments were carried out using synchrotron radiation at
beamline I311~\cite{Andersen1991} of the MAX radiation laboratory
(Lund, Sweden), the LEEM experiments with the LEEMIII instrument at
this beamline.

\begin{figure}
\begin{center}
\includegraphics[width=0.3\textwidth]{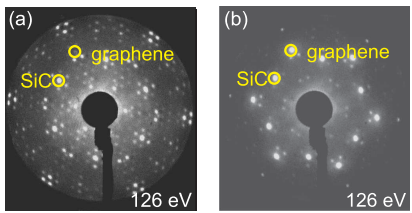}
\end{center}
\caption{LEED patterns at 126 eV for the {\sixroot}
reconstruction (zerolayer graphene) (a) before and (b) after
hydrogen intercalation. The first order diffraction spots are
indicated for SiC and graphene.} \label{leed}
\end{figure}

Figure \ref{leed} (a) and (b) display LEED patterns for the
{\sixroot} reconstructed buffer layer before and after hydrogen
treatment. For the pristine buffer layer (panel (a)) the LEED
pattern shows intense superstructure spots corresponding to the
pronounced atomic displacements in this layer due to the covalent
bonding to the SiC substrate~\cite{Riedl2007}. After hydrogen
treatment, the superstructure spots are strongly suppressed as
depicted in panel (b) which indicates much smaller atomic
displacements in the reconstructed layers which in turn suggests the
absence or weakening of the interlayer bonding. Similarly, for an
epitaxial monolayer the spots of the {\sixroot} superstructure
vanish upon hydrogen treatment (not shown). This is already a clear
indication of a geometrical decoupling of the interface layer from
the substrate.

\begin{figure}[h!]
\begin{center}
\includegraphics[width=0.48\textwidth]{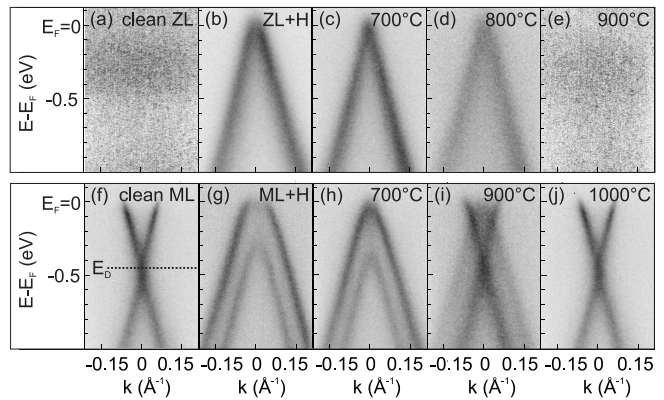}
\end{center}
\caption{Dispersion of the $\pi$-bands measured with
ARPES perpendicular to the $\bar{\Gamma}\bar{\textrm{K}}$ direction
of the graphene Brillouin zone for (a) an as-grown graphene
zerolayer (ZL) on SiC(0001), (b) after hydrogen treatment and (c-e)
subsequent annealing steps; (f) for an as-grown monolayer (ML), (g)
after hydrogen treatment and (h-j) subsequent annealing steps.}
\label{ups}
\end{figure}

Apart from this structural aspect, the hydrogen treatment has a
dramatic effect on the electronic structure of the samples. This is
shown in Figure \ref{ups} by ARPES measurements of
the valence band structure around the $\bar{\textrm{K}}$-point of
the graphene Brillouin zone. For a pristine zerolayer
no $\pi$-bands are observed as displayed in panel
(a). Only two very faint delocalized and smeared out states at
binding energies of around 0.1eV to 0.5eV and higher than 0.9eV are
visible. For a zerolayer sample after hydrogen treatment, quite
differently, the linear dispersing $\pi$-bands of monolayer graphene
appear, see panel (b). In addition, while as-grown
monolayer graphene is n-doped, so that its Fermi level
$\textrm{E}_\textrm{F}$ is located about 420 meV above the crossing
point of the $\pi$-bands (Dirac point, $\textrm{E}_\textrm{D}$), as
shown below (Fig. \ref{ups} (f)), this effect is
reversed after the hydrogen treatment and the sample is slightly
p-doped so that $\textrm{E}_\textrm{F}$ is shifted below
$\textrm{E}_\textrm{D}$ by $\approx$ 100 meV. The appearance of the
graphene type $\pi$-bands and the absence of n-doping corroborates
our working hypothesis that the covalently bound carbon layer is
decoupled from the substrate. Apparently, the hydrogen atoms migrate
under this layer, break the bonds between C and Si and bind to the
Si atoms as sketched in Fig. \ref{models} (a) and
(c). Correspondingly, the buffer layer is lifted and displays the
electronic properties of a quasi-free standing graphene monolayer.
Note, that outgassing the hydrogen treated sample at 400{\cgrad} as
carried out for the spectrum shown in Fig. \ref{ups}
(b) has no effect on the $\pi$-band structure. Yet,
after heating the sample up to 700{\cgrad} (panel
(c)) the slight p-doping vanishes and charge neutrality is
retrieved ($\textrm{E}_\textrm{F}$=$\textrm{E}_\textrm{D}$), so that
we tentatively attribute the p-doping effect to the presence of
chemisorbed species on the graphene surface and the subsequent
downshift of the band structure to their desorption \cite{Lee2008}.
At temperatures above 700{\cgrad} the $\pi$-bands progressively
weaken as indicated in panel (d). Since Si-H bonds
are known to break at temperatures just above 700{\cgrad}
\cite{Sieber2003}, this effect can be correlated to progressive
hydrogen desorption. Around 900{\cgrad} the hydrogen has completely
desorbed and the zerolayer structure is re-established as seen from
the absence of $\pi$-bands (Fig. \ref{ups}(e)) and
also from the LEED pattern which is similar again to the one shown
in Fig. \ref{leed} (a). Consistent results were
observed for monolayer epitaxial graphene, which turns into bilayer
graphene upon hydrogen treatment as previously
sketched in Fig. \ref{models} (c) and (d). The corresponding
bandstructure measured by ARPES before hydrogen
treatment and after hydrogen treatment plus subsequent outgassing at
400{\cgrad} is shown in Fig. \ref{ups} (f) and
(g). Again the hydrogen treated sample shows a
slight p doping which disappears after annealing to 700{\cgrad}
(panel (h)). For temperatures higher than
700{\cgrad} the intensity of the bilayer $\pi$-bands decreases while
the monolayer bands reappear (Fig. \ref{ups} (i) and
(j)). The hydrogen progressively desorbs until at 1000{\cgrad} the
original monolayer bandstructure is completely recovered (not
shown).

\begin{figure}
\begin{center}
\includegraphics[width=0.47\textwidth]{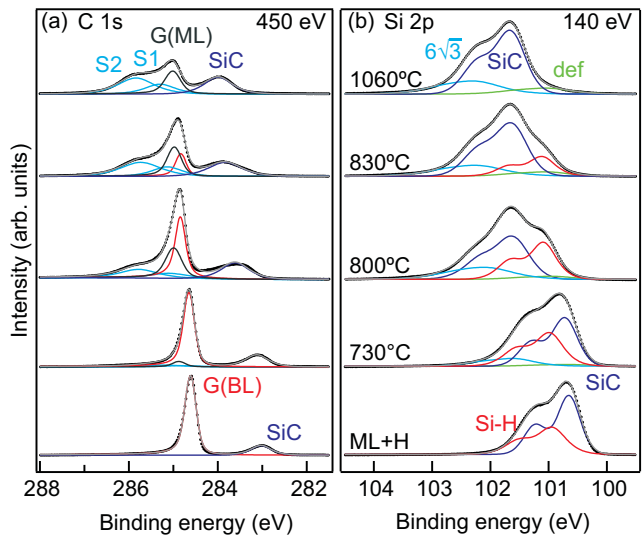}
\end{center}
\caption{C 1s (a) and Si 2p (b) core level spectra for a hydrogen
treated monolayer graphene sample (bottom spectra) and the same
sample annealed at increasing temperatures. The experimental data
are displayed in black dots. Different components, accordingly
labeled in the spectra, are fitted into the C 1s and Si 2p regions
by a line shape analysis. The gray solid line is the envelope of the
fitted components.} \label{xps}
\end{figure}

\begin{figure}[h!]
\begin{center}
\includegraphics[width=0.4\textwidth]{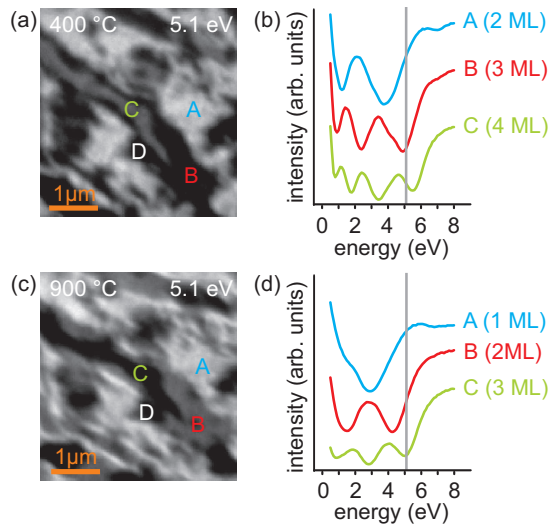}
\end{center}
\caption{4${\times}4$ \textrm{${\mu}m^{2}$} LEEM micrographs
recorded with an electron energy of 5.1 eV for the same area of (a)
a hydrogen-treated graphene sample after outgassing at 400{\cgrad}
and (c) annealed at 900{\cgrad}. Representative regions are labeled
A, B, C, D. The electron reflectivity spectra obtained for the
regions A, B, and C are plotted in panels (b) and (d), respectively,
labeled with the number of graphene monolayers (ML). } \label{leem}
\end{figure}

To corroborate the structural models sketched in Fig.
\ref{models}, a detailed analysis of the chemical
bonds was conducted by using CLPES. Since the resulting picture is
very similar for zerolayer and monolayer graphene only the monolayer
results will be discussed. Si 2p and C 1s core level spectra were
measured for the hydrogen-treated monolayer sample annealed at
different temperatures, as depicted in Fig.
\ref{xps}. Different components contributing to the
spectra were decomposed by a curve fitting procedure~\cite{fitting}.
The depth position of the corresponding species within the surface
was identified by varying the incident photon energy and thus
changing the surface sensitivity; the energies shown in Fig.
\ref{xps} are 140 eV and 450 eV for the Si 2p and C
1s spectra, respectively. The experimental data points are displayed
in black dots. The gray solid line is the envelope of the fitted
components. The bottom curve in Fig. \ref{xps} (a)
shows the C1s spectrum measured after outgassing the
hydrogen-treated sample at a temperature around 600{\cgrad}. The
dominant peak at 284.6 eV is the graphene related component (red
line) while the broader, less intense peak at 283.0 eV is the SiC
(bulk) related component (dark blue line). We emphasize the complete
absence of components related to the {\sixroot}
reconstruction~\cite{Emtsev2008}, which clearly identifies the
surface as a quasi-free standing epitaxial graphene layer (bilayer
in this case). At annealing temperatures higher than 700{\cgrad} the
hydrogen starts to desorb, as indicated by the appearance of the
interface components S1 and S2, marked with light blue lines. They
result from the carbon atoms in the {\sixroot} structure
\cite{Emtsev2008}. The hydrogen desorption implicates the appearance
of a second graphene related peak (black line), representing those
patches where the hydrogen has left. For an annealing temperature of
830{\cgrad} this monolayer component becomes more significant than
the one resulting from the quasi-free bilayer patches. After
annealing at 1000{\cgrad} the hydrogen is completely desorbed: the C
1s spectrum acquires the shape typically obtained for epitaxial
monolayer graphene. The difference in binding energy location
between the monolayer and decoupled bilayer components is about 0.4
eV, which perfectly agrees with the shift of the Fermi level
measured via ARPES. The total shift of the SiC component is 1 eV,
which confirms that on the SiC surface hydrogen bonds are present
which cause a respective band bending.

Further evidence of the existence of Si-H bonds is brought by the Si
2p data. The Si 2p spectrum obtained after initial outgassing
(bottom curve in Fig. \ref{xps} (b)) consists of two spin-orbit
split doublets. The binding energies are given with respect to the
Si 2p$_{3/2}$ component. According to the surface sensitivity
variations and in agreement with Ref. \cite{Sieber2003}, the
dominant peak at 100.6 eV (dark blue line) can be assigned to the
bulk component and the one at 100.9 eV (red line) to Si-H bonds.
After annealing at 730{\cgrad} the Si 2p spectrum can be accurately
fitted only by introducing two additional components: the one at
higher binding energy (light blue line) is attributed to the Si
atoms bonded to the {\sixroot} reconstructed overlayer, the small
one at lower binding energy (green line) to surface defects. These
components increase in intensity for increasing annealing
temperatures while the Si-H component gradually vanishes and
completely disappears after annealing at 1000{\cgrad}. The total
shift observed for the Si 2p bulk component amounts to 1 eV in
agreement with the C 1s bulk peak.

The effect of hydrogen intercalation on the graphene structure can
be analysed with spatial resolution using LEEM, a method that can
identify the number of graphene layers on SiC by the number of dips
in the electron reflectivity spectra between 0 and 8
eV~\cite{Hibino2008}. In Fig. \ref{leem}, LEEM micrographs are shown
for an electron energy of 5.1 eV measured in the same area of the
sample with (panel (a)) and without (panel (c)) intercalated
hydrogen. At this energy, regions of different graphene thickness
can be distinguished by the reflected intensity. The electron
reflectivity spectra for the different surface domains A, B and C as
labeled in panel (a) are plotted in panel (b). The number of dips in
the spectra identifies region A, B and C as bi-, tri-, and four
layer graphene. After desorbing the hydrogen through an annealing
step at 900{\cgrad}, the spatial distribution of these domains does
not change as shown in panel (c). However, their LEEM intensity
changes and the reflectivity spectra as plotted in panel (d)
identify a complete transformation of (n+1)-layer thick areas into
(n)-layer thick areas (n=1,2,3). Note that the region labeled D in
Fig. \ref{leem} displays the same intensity before and after
desorption of the hydrogen (and a flat reflectivity spectrum) and is
attributed to surface defects, e.g.\ from residual polishing damage.

The question arises how the hydrogen migrates both below the
interface layer and even through several graphene layers. Recent
experimental reports on hydrogen on graphene did not show any
evidence for hydrogen penetration through graphene \cite{Elias2009,
Bostwick2009, Guisinger2009,Balog2009}. However, in contrast to
these experiments we use molecular hydrogen at atmospheric pressures
and our graphene samples were annealed up to 1000{\cgrad} which
might facilitate a reactive passage of the hydrogen through the
graphene lattice. Another possibility is that the hydrogen
intercalation starts at grain boundaries or defects on the surface.
On a final note, we point out that the hydrogen intercalated samples
are extremely stable in ambient atmosphere, at least for several
months. Furthermore the hydrogen passivation and
desorption can be repeated several times without notable changes in
the sample quality.

In summary, we have demonstrated that hydrogen can migrate through
epitaxial graphene and the interface layer, bind to the Si atoms of
the SiC(0001) surface and decouple epitaxial graphene from its
substrate. n-layer graphene films transform into (n+1)-layer
graphene films (n=0,1,2,3). The intercalation opens up the
possibility to produce quasi-free standing epitaxial graphene on
large SiC wafers. The hydrogen passivates the underlying SiC
substrate similar to the case of bare SiC
surfaces~~\cite{Tsuchida1999, Seyller2004, Coletti2008}. The
intercalated hydrogen is sustained in ambient conditions and stable
up to 700{\cgrad}. The intercalation process is technologically well
adapted and represents a highly promising route towards epitaxial
graphene based nanoelectronics.

Support by the EC through the Access to Research Infrastructure
Action is greatfully acknowledged. C.C. and T.I. acknowledge the
Alexander von Humboldt research fellowship for financial support.

%\newpage


\begin{thebibliography}{00}
\bibitem{fitting} A linear background was subtracted from the Si 2p
data and a Shirley background from the C 1s data. The C 1s graphene
related peak was fitted using a Doniach-Sunjic profile to account
for the metallic behavior, all the other fitted lines are an
approximation of the Voigt function.
\end{thebibliography}
\end{document}